\documentclass[11pt,a4paper]{article}

\usepackage[margin=2.2cm]{geometry}
\usepackage[T1]{fontenc}
\usepackage{lmodern}
\usepackage{amsmath}
\usepackage[round]{natbib}
\usepackage{booktabs}
\usepackage{enumitem}
\usepackage{authblk}
\usepackage[colorlinks=true,linkcolor=blue,citecolor=blue,urlcolor=blue]{hyperref}
\usepackage{xcolor}
\usepackage{orcidlink}
\newcommand{\Hnot}{\ensuremath{H_0}}
\newcommand{\kms}{\ensuremath{\mathrm{km\,s^{-1}\,Mpc^{-1}}}}

\newcommand{\affiliation}[1]{\footnotesize #1}

\newcommand{\desy}{\affiliation Deutsches Elektronen-Synchrotron DESY,
Platanenallee 6, 15738 Zeuthen, Germany}

\newcommand{\HUBerlin}{\affiliation{Institut f\"ur Physik, Humboldt-Universit\"at zu Berlin, Newtonstr. 15, 12489 Berlin, Germany}}

\newcommand{\Lyon}{\affiliation{Universit\'e Lyon 1, CNRS, IP2I Lyon, UMR 5822, Villeurbanne, France}}

\newcommand{\ICEN}{\affiliation Instituto de Ciencias Exactas y Naturales (ICEN), Universidad Arturo Prat, Chile}

\newcommand{\lancaster}{\affiliation School or Physics and Astronomy, Lancaster University, Lancaster, LA1 4YB, UK}

\newcommand{\yonsei}{\affiliation Department of Astronomy \& Center for Galaxy Evolution Research, Yonsei University, Seoul 03722, Republic of Korea}

\newcommand{\OKC}{\affiliation{Department of Physics, Oskar Klein Centre, Stockholm University, SE-106 91, Stockholm, Sweden}}

\newcommand{\IoAKavli}{\affiliation{
Institute of Astronomy and Kavli Institute for Cosmology, University of Cambridge, Madingley Road, Cambridge, CB3 0HA, UK}}

\newcommand{\Southampton}{\affiliation{Department of Physics and Astronomy, University of Southampton, Southampton SO17 1BJ, UK}}

\newcommand{\Birmingham}{\affiliation{School of Physics \& Astronomy and Institute for Gravitational Wave Astronomy, University of Birmingham, Birmingham, B15 2TT, UK}}

\newcommand{\ICE}{\affiliation{Institute of Space Sciences (ICE, CSIC), Campus UAB, Carrer de Can Magrans, s/n, E-08193 Barcelona, Spain}}
\newcommand{\IEEC}{\affiliation{Institut d’Estudis Espacials de Catalunya (IEEC), E-08034 Barcelona, Spain}}

\newcommand{\UDublin}{\affiliation{School of Physics, Trinity College Dublin, The University of Dublin, Dublin 2, Ireland}}

\newcommand{\LPNHE}{\affiliation{LPNHE, (CNRS/IN2P3, Sorbonne Universit\'{e}, Universit\'{e} Paris Cit\'{e}), Laboratoire de Physique Nucl\'{e}aire et de Hautes \'{E}nergies, 75005, Paris, France}}

\title{Comment on the recent claims of $H_0$ measurements with the ZTF DR2 SN~Ia sample (arXiv:2610.36051)}
\author[1]{M. Rigault\thanks{\texttt{m.rigault@ip2i.in2p3.fr}}
\orcidlink{0000-0002-8121-2560}}
\author[2]{B. Popovic\thanks{\texttt{b.a.popovic@soton.ac.uk}}
\orcidlink{0000-0002-8121-2560}}
\author[3]{A. Goobar\thanks{\texttt{ariel@fysik.su.se}}\orcidlink{0000-0002-4163-4996}}
\author[4]{S. Dhawan\thanks{\texttt{s.dhawan@bham.ac.uk}}
\orcidlink{0000-0002-2376-6979}}
\author[1]{M. Aubert\orcidlink{0009-0002-7667-8814}}
\author[5]{M.~Betoule\orcidlink{0000-0003-0804-836X}}
\author[6]{U. Burgaz\orcidlink{0000-0003-0126-3999}}
\author[7]{G. Dimitriadis\orcidlink{0000-0001-9494-179X}}
\author[8, 9]{L. Galbany\orcidlink{0000-0002-1296-6887}}
\author[10]{M. Ginolin\orcidlink{0009-0004-5311-9301}}
\author[3]{A.~Gilles-Lordet\orcidlink{0009-0002-4464-8063}}
\author[11]{Y.-L. Kim\orcidlink{0000-0002-1031-0796}}
\author[12, 13-]{M. Kowalski\orcidlink{0000-0001-8594-8666}}
\author[8, 9]{T. E. Müller-Bravo\orcidlink{0000-0003-3939-7167}}
\author[12]{J. Nordin\orcidlink{0000-0001-8342-6274}}
\author[5]{N.~Regnault\orcidlink{0000-0001-7029-7901}}
\author[7]{M. Smith\orcidlink{0000-0002-3321-1432}}
\author[3]{J. Sollerman\orcidlink{0000-0003-1546-6615}}
\author[6]{J. H. Terwel\orcidlink{0000-0001-9834-3439}}
\author[4]{A. Townsend\orcidlink{0000-0001-6343-3362}}
\author{on behalf of the ZTF SN Ia DR2 Collaboration}
\affil[1]{\Lyon}
\affil[2]{\Southampton}
\affil[3]{\OKC}
\affil[4]{\Birmingham}
\affil[5]{\LPNHE}
\affil[6]{\UDublin}
\affil[7]{\lancaster}
\affil[8]{\ICE}
\affil[9]{\IEEC}
\affil[10]{\IoAKavli}
\affil[11]{\yonsei}
\affil[12]{\desy}
\affil[13]{\HUBerlin}
\affil[14]{\ICEN}

\date{\today}

\begin{document}
\maketitle

\begin{abstract}
\citet{karchev} derived a value of $H_0$ by combining the ZTF SN Ia DR2 sample with SH0ES calibrator Type Ia supernovae (SNe Ia; \citealt{riess2022}), used to anchor the SN Ia absolute luminosity.
The ZTF SN Ia DR2 dataset \citep{rigault25} is the largest ($\mathrm{O}(3000)$) SN Ia sample to date obtained with a single telescope. However, as emphasized in the data release, the absolute photometric calibration of the survey has not yet been established to the precision required for cosmological inference.
The relative photometric zeropoint offset between the SH0ES and ZTF SN Ia DR2 samples, which is fully degenerate with $H_0$, is therefore unknown and cannot be determined from publicly available data.
The photometric corrections introduced in \citet{karchev} do not address this global calibration offset.
Consequently, their reported $H_0$ determination is not supported by the available calibration information.
We nonetheless share the excitement of pursuing state-of-the-art cosmology with ZTF and are actively working toward the release of a fully calibrated dataset.
\end{abstract}

\section{Introduction and summary of the problem}

Type~Ia supernovae (SNe~Ia) are exploding stars with standardizable peak luminosity, so their apparent  brightness tells us their distance.
The Hubble constant \Hnot{} is measured directly by comparing two groups of SNe Ia: a few nearby ``calibrator'' SNe whose distances are known independently (e.g., from Cepheid stars, themselves anchored by geometric distances such as parallaxes), and many more comparatively distant ($z \sim 0.02 - 0.1$) SNe~Ia that trace the expansion of the Universe. \Hnot{} follows from how bright the distant SNe~Ia are relative to the calibrator ones. The method therefore only works if both groups are measured on {\em exactly the same brightness scale}. 

Interestingly, these direct measurements are in disagreement at the $\geq5\sigma$ level with the prediction of the standard model of cosmology anchored by the cosmic microwave background \citep[e.g.,][]{riess2022, planck2020}. This issue, known as the Hubble Tension, represents one of the most exciting fronts in cosmology today. 

In that context, the Zwicky Transient Facility (ZTF) SN Ia sample provides a novel low-redshift ($z\leq0.1$) data set of nearby SNe Ia that can be used to measure $H_0$ \citep{2022ApJ...934..185D} and time-varying dark energy $w_0/w_a$ \citep{rigault25}; but to accurately measure cosmology, equally-accurate calibration within ZTF and across other samples is required. However, as highlighted in \cite{rigault25}, the zero point (ZP) calibration of the ZTF SN Ia DR2 sample (i.e., the brightness scale) is unknown and cannot be accurately measured with the current public release, for reasons that are detailed in this letter.

\citet{karchev} attempt to combine the ZTF SN Ia DR2 sample with the historic SH0ES sample of calibrator SNe Ia, assuming both to be in the same photometric system. This assumption is unsupported by the available calibration data. Since the unknown relative calibration offset is exactly degenerate with $\Hnot$, the published result and associated claims cannot be regarded as a valid measurement of $\Hnot$ with the quoted uncertainty.

\vspace{-0.4cm}
\section{Requirements to Measure \Hnot}

Measurement of \Hnot~through the distance ladder technique requires having all distance indicators in the same photometric system, i.e., that the magnitude definition is consistent across surveys and samples. When comparing the calibrator SN Ia sample (which constrains the SN luminosity) and the Hubble flow sample (which constrains $H_0$ given a known SN luminosity), any unaccounted for magnitude offset "$\Delta(mag)$" between the samples directly biases the derivation of $H_0$ by "$\Delta({H_0})$", such that
$$
\Delta({H_0})/H_0 = -\frac{\ln{10}}{5}  \Delta(mag).
$$
Consequently, a calibration ZP offset between the calibrator sample (SH0ES) and the Hubble flow sample (ZTF) of, say 5\% (0.05 mag), leads to a bias on $H_0$ of $\sim1.7$~\kms{}.

 Highlighted in \cite{rigault25} and further discussed in \cite{lacroix25}, the published ZTF SN Ia DR2 sample suffers from clear calibration and non-linearity issues. Most notably here, the "DR2" sample is not tied with sufficient accuracy to any photometric calibration system, and there are no data products publicly available that would allow ZTF to be combined with other photometric systems; hence the pure-ZTF $H_0$  demonstration analysis presented in \cite{2022ApJ...934..185D}.
 The above issues have not been addressed by \cite{karchev}, resulting in a value of \Hnot{} that is not supported by the data. 
\vspace{-0.4cm}
\section{The Zero Point system}

The SH0ES calibrator sample is a compilation of several different telescopes and photometric systems anchored to a common reference one: Pan-STARRS \citep[PS1, see details in][]{fragilistic}. 

In the context of the ZTF SN Ia DR2 sample, \cite{rigault25} released SN Ia lightcurves derived from template-subtracted forced-photometry (FP), comparing images with the SN to images without. In the official ZTF pipeline \citep{masci2019}, images are anchored to PS1 using a linear colour correction and assuming a homogeneous filter profile as provided by the manufacturer. Under these assumptions, \cite{masci2019} suggest that ZTF images are calibrated to the 2 to 3\% level. Subsequent ZTF data releases have unveiled significant additional ZP calibration issues,\footnote{\url{http://nesssi.cacr.caltech.edu/ZTF/Web/Calib.html}} in addition to these presented in \cite{lacroix25}. 
Some of these calibration issues have been mitigated at the star catalog level on ZTF releases, but none of these corrections were applied to the science images, and are consequently not reflected in the released ZTF SN Ia DR2 lightcurves.

We briefly summarize key-steps of the photometric calibration chain, highlighting what is missing to calibrate the ZTF SN Ia DR2 dataset, and consequently why the value of $H_0$ inferred by \cite{karchev} is not a valid cosmological measurement.

\begin{enumerate}[leftmargin=*,itemsep=2pt]
  \item \textbf{Stellar Magnitudes.} 
  Combining photometric systems requires observing the same field stars with different telescopes. If one knows the filter bandpasses of two systems and their relative ZP, one can combine these systems into a shared photometric calibration, constraining the ZP and effective wavelength of the filters \citep[see details in e.g.][]{fragilistic, dovekie}. 
  
  No such stellar magnitudes were released as part of the ZTF SN Ia DR2. Consequently, any measurement of $H_0$ performed with the public ZTF DR2 SN Ia currently carries with it an unknown floating constant, lacking the data to place ZTF and the calibrator sample in the same magnitude system, e.g. that of PS1. This central issue is briefly mentioned in \cite{karchev} but then disregarded for their results and conclusions.

In addition, \textit{in-situ} measurements have been performed on ZTF filters using the Collimated Beam Projector \citep{mondrik2023, souverin2025}, and we now know that the public ZTF bandpasses are, on average, incorrect on the order of a few nm, and that bandpasses vary significantly across the focal plane. These updated measurements, necessary to ensure a common photometric system, are not yet publicly available. 
    
\item \textbf{Supernova Magnitudes.}
  The SN magnitudes, tied to those of the field stars, must also be determined. Unlike field stars, which are isolated and can be measured relatively simply with aperture photometry (AP), the SN Ia flux signal must be separated from that of their hosts using PSF modeling. This can be done with either forced-photometry on difference images (FP) or using scene modeling photometry (SMP). 
  The ZTF SN Ia DR2 lightcurves are obtained using FP; we presented an alternative flux extraction using SMP for comparison in \cite{lacroix25}.
  This unveiled a magnitude difference between the FP and SMP-derived lightcurves in ZTF, originally reported at the 90 mmag level \citep{lacroix25},  which has since been revised to $\sim60$ mmag in the accepted version, in closer agreement with the independent comparison published by \cite{marlin2026}. This difference in lightcurve processing alone shifts $H_0$ by $\sim1.7$~\kms{}.
  We emphasize nonetheless that any update on the FP-SMP offset does not account for the different photometric systems between ZTF and SH0ES highlighted in the first point, nor does this SMP-FP offset have well-understood uncertainties, thereby making an accurate modeling of the calibration errors untractable with published data.

  \item \textbf{PSF non-linearity } To close the calibration loop, the stellar PSF magnitudes (here, ZTF SMP output) must be compared to star AP measurements used to anchor the calibration system ; hence connecting all parts of the calibration chain.
  Doing so, we discovered a significant PSF non-linearity, dubbed `the pocket-effect', that results in a nonlinear flux-dependent bias \citep{lacroix25,rigault25}. This effect is strongly CCD dependent and thus varies across the camera footprint. The pocket effect cannot be patched \textit{a posteriori} as done by \cite{karchev} as it varies with survey time, the camera footprint, and impacts both the shape and the centroid of the resulting PSF. 
  We further highlight that a significant fraction of the ZTF CCDs suffer from pocket effect up to 10x larger than that assumed by \cite{karchev}, illustrating the inaccuracy of the patching approach employed by the authors. Finally, additional non-nonlinearities exist in our detector, such as the brighter-fatter effect, which affect the PSF flux calibration at the (few) percent(s) level. This hence has significant consequences on the derivation of cosmological parameters.
\end{enumerate}

These are some illustrations of publicly known issues that led us to explicitly state that the ZTF SN Ia DR2 cannot be used to derive cosmology \citep{rigault25}.
Altogether, the calibration offset between the ZTF SN Ia DR2 sample and the calibrator SH0ES sample is unknown and could be as large as 0.1 mag, corresponding to a systematic offset of 3.5~\kms{} on $H_0$.

We consequently consider the results of \cite{karchev}, such as the stated $2.7\sigma$ tension with \textit{Planck}, the claim of an independent confirmation of the Hubble Tension and the role of SN Ia in this context, to be unsupported by the public ZTF SN Ia DR2 data. Determining \Hnot~ with ZTF requires additional data products which are not yet released: magnitudes of the calibrator field stars, pocket- and nonlinearity-corrected photometry, the complete passbands, and the updated SMP lightcurves.

\vspace{-0.4cm}
\section{Summary and Outlook}
The ZTF SN Ia DR2 data are internally self-consistent for studies that do not need the associated ZP. Any pure-ZTF analyses or derived products, such as SN Ia population studies, etc. are consistent and unaffected by the issues highlighted here; it is only when combining ZTF with other surveys that these problems arise. 
We share the excitement of producing state-of-the-art cosmology with ZTF SN Ia data. For that endeavor, the ZTF Cosmology science working group has been working on a proper calibration of the survey. The final release of the ZTF SNe Ia will include the associated data products that will enable anyone to replicate or alter the analyses.

As it stands, the \cite{karchev} manuscript is in error and does not represent a measurement of $\Hnot$. Would the authors still choose to publish their analysis, the limitations outlined here should be made clearly explicit.



\begin{thebibliography}{9}
\begin{small}
\bibitem[Brout et al.(2022a)]{fragilistic} Brout, D., Taylor, G., Scolnic, D., et al.\ 2022, The Astrophysical Journal, 938, 111. doi:10.3847/1538-4357/ac8bcc.

\bibitem[\protect\citeauthoryear{Dhawan et al.}{2022}]{2022ApJ...934..185D} Dhawan S., Goobar A., Johansson J., Jang I.~S., Rigault M., Harvey L., Maguire K., et al., 2022, ApJ, 934, 185. doi:10.3847/1538-4357/ac7ceb
 
\bibitem[Karchev et al.(2026)]{karchev} Karchev, K., Giunta, M., Jimenez, R., Trotta, R., \& Verde, L. 2026, arXiv:2610.36051

\bibitem[Lacroix et al.(2025)]{lacroix25} Lacroix, L., Regnault, N., de Jaeger, T., et al.\ 2025, arXiv:2509.04073. doi:10.48550/arXiv.2509.04073

\bibitem[Marlin et al.(2026)]{marlin2026} Marlin, E.~G., Murakami, Y.~S., Brout, D., et al.\ 2026, ApJ, 1004, 2, 173. doi:10.3847/1538-4357/ae69cc

\bibitem[Masci et al.(2019)]{masci2019} Masci, F.~J., Laher, R.~R., Rusholme, B., et al.\ 2019, Publications of the Astronomical Society of the Pacific, 131, 018003. doi:10.1088/1538-3873/aae8ac.

\bibitem[Mondrik et al.(2023)]{mondrik2023} Mondrik, N., Coughlin, M., Betoule, M., et al.\ 2023, Publications of the Astronomical Society of the Pacific, 135, 035001. doi:10.1088/1538-3873/acbe1c.

\bibitem[Planck Collaboration et al.(2020)]{planck2020} Planck Collaboration, Aghanim, N., Akrami, Y., et al.\ 2020, A\&A, 641, A6. doi:10.1051/0004-6361/201833910


\bibitem[Popovic et al.(2026)]{dovekie} Popovic, B., Kenworthy, W.~D., Ginolin, M., et al.\ 2026, Astronomy and Astrophysics, 712, A131. doi:10.1051/0004-6361/202555963.

\bibitem[Riess et al.(2022)]{riess2022} Riess, A.~G., Yuan, W., Macri, L.~M., et al.\ 2022, The Astrophysical Journal Letters, 934, L7. doi:10.3847/2041-8213/ac5c5b.


\bibitem[Rigault et al.(2025)]{rigault25} Rigault, M., Smith, M., Goobar, A., et al.\ 2025, Astronomy and Astrophysics, 694, A1. doi:10.1051/0004-6361/202450388.

\bibitem[Souverin et al.(2025)]{souverin2025} Souverin, T., Neveu, J., Betoule, M., et al.\ 2025, RAS Techniques and Instruments, 4, rzaf010. doi:10.1093/rasti/rzaf010

\end{small}
\end{thebibliography}
\end{document}